\documentclass[prc,aps,preprint]{revtex4}  
\usepackage{graphicx} 
\begin{document} 
\draft
\title{ Short-range correlations in the deuteron: chiral effective field theory, meson-exchange, and  
phenomenology} 
\author{            
Francesca Sammarruca      }                                                                                
\affiliation{ Physics Department, University of Idaho, Moscow, ID 83844-0903, U.S.A. 
}
\date{\today} 
\begin{abstract}
We study high-momentum distributions and short-range correlation probabilities in the deuteron with 
a variety of modern potentials based on chiral effective field theory up to fifth order in the chiral expansion.   
 We also consider some 
conventional 
(meson-exchange and phenomenological) interactions. We examine our predictions in the context 
of short-range correlation probabilities as extracted from analyses of inclusive electron scattering 
data and discuss whether modern interactions can be reconciled with the latter. 
\end{abstract}
\maketitle

\section{Introduction} 
\label{Intro} 

High-momentum components in the nuclear wave function and in momentum distributions
are a reflection of short-range correlations (SRC) in nuclei. 
The presence of high-momentum components is mainly due to the repulsive short-range central force and the
tensor force. 

Although high-quality meson-theoretic interactions continue to be employed in contemporary calculations 
of nuclear structure and reactions, since the 1990's there exists 
a general understanding that chiral effective field theory (EFT) \cite{Wei68,Wein79} is a superior
 framework. First, chiral EFT
has a firm connection with quantum chromodynamics (QCD) through the symmetries of low-energy QCD.             
Second, it allows for a systematic expansion which makes possible a 
quantification of the theoretical error.                                          
At each order of chiral perturbation theory ($\chi$PT), the uncertainty associated 
with a particular prediction can be controlled and quantified. For these reasons, nuclear chiral
effective theory is becoming 
increasingly popular as a model-independent approach.                                  

The NN potentials constructed within chiral EFT are generally softer than ``conventional" potentials,
which makes them computationally more amenable to nuclear structure calculations. 
Also, potentials with a low resolution scale, obtained through a        
unitary transformation (that is, through 
renormalization group (RG) methods~\cite{RG}) applied to a ``harder" interaction, are very popular for 
many-body calculations. 
The resulting ``low-k" potentials are equivalent to the original ones for all physical purposes, although 
essentially void of high momentum components.

On the experimental side, 
inclusive electron scattering measurements at high momentum transfer, on both light and heavy nuclei,
have been analyzed with the purpose of extracting information on short-range correlations~\cite{CLAS,src,Pia+}.
In a suitable range of $Q^2$ and $x_B$, the cross section is factorized in order to single out the probability of a nucleon
to be involved in SRC, either two-body or three-body. 
When extended to nuclear matter, this probability is equivalent to the ``wound integral", which measures the amount 
of correlations in the so-called defect function~\cite{FS14}. 
Information about two-body correlations can also be obtained in coincidence experiments involving knock-out of a nucleon
pair with protons~\cite{Tang} or electrons~\cite{Korover,Shneor,Subedi,Baghda}. 

Nuclear scaling and the plateaus seen in inclusive scattering cross section ratios~\cite{CLAS} 
are due to the dominance of SRC for momenta above approximately 2 fm$^{-1}$. In the same region,
the momentum distribution in a nucleus relative to the one in the deuteron becomes almost flat, so that those distributions
simply scale with $A$.              

The discussion around some of these measurements which is presently going on in the literature is quite intriguing.  
The probabilities mentioned above are a manifestation of the off-shell nature 
of the potential, which cannot be 
determined uniquely from NN elastic data and is not an observable. Interactions may differ dramatically 
in their off-shell behavior while remaining phase-equivalent.                                           
The most striking example is provided by the RG-evolved potentials we mentioned above, where 
the high-momentum structures of the original and the RG-evolved potential are obviously not the same.    

Naturally, 
a low-resolution scale will impact the ability to resolve high momentum regions.    
With regard to this point,        
it has been noted that, if a unitary transformation is applied to both wave functions and 
operators, one regains the invariance of the cross section, as one should~\cite{Furn13}, thus attaining a 
consistent description of short- and long-range physics. This has been addressed recently by Neff {\it et al.},
who show how the short-range information can be recovered by transforming the density operators~\cite{Neff}.

On the other hand, chiral potentials such as those developed in Refs.~\cite{EM03,ME11,chinn5} are not low-momentum interactions in the sense of a $V_{low-k}$. In this paper, we examine those from the point of view of SRC. 
To broaden the discussion, we start with an analysis of SRC and conventional (that is, non-chiral) NN potentials, including high-precision potentials from the 1990's as well as a phenomenological one.                         
We then move to a similar analysis with chiral interactions. 

 We take the deuteron as our sample system. We recall that 
the high-momentum part of the momentum distribution shows 
similar features in nuclei with $A$=2 to 40~\cite{Alv13}. Thus, the deuteron offers representative features. 
Furthermore, deuteron SRC probabilities are a crucial element in the estimation of SRC
probabilities in heavier nuclei as obtained in Ref.~\cite{CLAS}. 

We calculate the momentum distribution and the probability of SRC up to 5th order
of chiral effective theory. Working with the $A$=2 system, we can go to any order        
of chiral EFT where NN potentials are available, without the need to worry about the corresponding three-nucleon forces (3NF).
Although 3NF at N$^3$LO and N$^4$LO have been worked out~\cite{3nf1,3nf2,3nf3}, their application in few- and many-body
systems still presents considerable challenges and requires unavoidable omissions/approximations.
Calculations in the deuteron are free of those and thus well-controlled. 
We address cutoff dependence and order-by-order
convergence from lowest to 5th order of the chiral expansion.       

Some of the questions we wish to address are: 
To which extent are modern, {\it non-phenomenological} interactions (chiral or not) consistent 
with the information as extracted from $A(e,e')X$ measurements?                                          
What do we learn, on fundamental grounds, from the answer to this question? 
Are there characteristic differences among particular families of potentials from which we can
obtain physical insight (beyond phenomenological observations)? 

Our findings and conclusions are summarized in Section~\ref{Concl}.

\section{High-momentum distribution in the deuteron} 
\subsection{Meson theory and phenomenology} 
\label{IIA} 

We begin with a step back into the 1990's by considering three members 
of the ``high-precision" family of NN potentials, namely CD-Bonn~\cite{CD}, 
Nijmegen II~\cite{nij}, and AV18~\cite{av18}. 
 The respective momentum distributions in the deuteron, with focus on high-momentum components,
are shown in Fig.~1.
$\rho(k)$ is the Fourier transform squared of the coordinate space wave function.                         
There are noticeable differences between the (softer) predictions from CD-Bonn and those from the other
two potentials, which are essentially indistinguishable. 

\begin{figure}[!t] 
\centering         
\vspace*{1.2cm}
\hspace*{0.5cm}
\scalebox{0.4}{\includegraphics{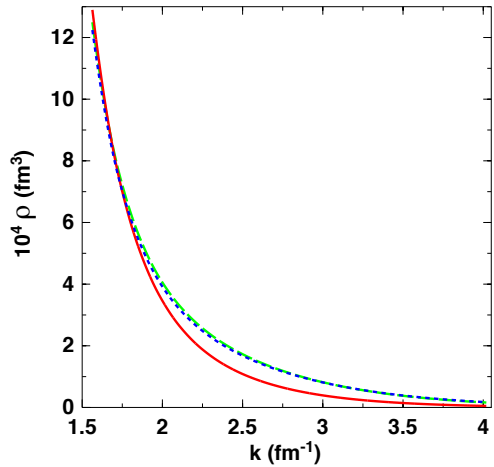}}
\vspace*{-2.5cm}
\caption{(Color online)                                                                         
High momentum distributions in the deuteron as predicted by: CD-Bonn (solid red); Nijmegen II (dotted blue); and
AV18 (dashed green). 
} 
\label{FIG1}
\end{figure}

\begin{table}                
\centering \caption                                                    
{Probabilities of SRC and $D$-state probabilities for the potentials considered in Fig.~1. 
} 
\vspace{5mm}
\begin{tabular}{|c|c|c|}
\hline
Model & $a_{2N}(d)$ & $P_D$ \\
\hline     
\hline     
 CD-Bonn &  0.032 & 0.0485  \\
   &     &   \\ 
 AV18 &  0.042& 0.0578  \\
  &      &    \\ 
  Nijmegen II &  0.041& 0.0565  \\
\hline
\end{tabular}
\label{tab1} 
\end{table}

To see how these differences carry into the probability of SRC, we 
follow Ref.~\cite{Alv13} and define the probability of SRC in the deuteron as 
\begin{equation}
a_{2N}(d) = 4 \pi \int _{k_{min}}^{\infty} \rho(k)k^2\;dk \; , 
\label{psrc} 
\end{equation} 
where $k_{min}$ is taken to be 1.4 fm$^{-1}$ (276 MeV). This definition was adopted in Ref.~\cite{CLAS}, where the 
choice of the lower integration limit is suggested by the onset of scaling of the cross section, which
should signal the dominance of scattering from a strongly correlated nucleon.
In Ref.~\cite{CLAS}, the ratio of the per-nucleon probability of two nucleon (2N) SRC in a nucleus relative to $^3$He is argued to be equal to the ratio of the inclusive electron scattering cross sections in the appropriate scaling region. 
The absolute per-nucleon probability in a nucleus can then be deduced if 
the absolute per-nucleon probability in $^3$He is known. The latter is the product of the absolute per-nucleon
probability in the deuteron, stated as 0.041$\pm$ 0.008 in Ref.~\cite{CLAS2}, and the relative probability of 2N SRC in 
$^3$He relative to the deuteron.                                                                                
Namely,  
\begin{equation}
a_{2N}(A)=a_{2N}(A/^3He)a_{2N}(^3He) \; \; \; \; \mbox{and} \; \; \; \; a_{2N}(^3He)=a_{2N}(^3He/d)a_{2N}(d) \;. 
\end{equation}
We note that the values for the deuteron and the ratio of $^3$He to deuteron 
contain some model dependence from theoretical calculations (see Ref.~\cite{CLAS} and Refs.~[2,6,15,16] therein), likely
to propagate in the predictions for heavier nuclei. 

In Table~\ref{tab1}, we show the probability as defined in Eq.~(\ref{psrc}) for the interactions used in Fig.~1. 
As an additional, related information, we also show the corresponding $D$-state probabilities. 
As expected in light of Fig.~\ref{FIG1}, 
there is a significant difference between CD-Bonn and the other two cases, with 
the AV18 and Nijmegen II predictions closer to the value used in the analysis from Ref.~\cite{CLAS2}.            
                  
The differences noted above are due to the 
non-local nature of CD-Bonn, which adopts fully relativistic momentum-space expressions for the one-pion-exchange.
More precisely, the off-shell nature of CD-Bonn is based upon the relativistic
Feynman amplitudes for meson exchange.                                                                            
This determines well-founded non-localities in the tensor force, whereas Nijmegen II and AV18 make use of the non-relativistic, static
one-pion-exchange which generates a local tensor force.
The characteristically softer nature of a relativistic momentum-space potential reflected in Table~\ref{tab1} is
a desirable feature for the purpose of applications in nuclear structure. 

We will come back to this point for a more complete discussion after the next section.         

\subsection{Interactions based on chiral EFT} 
\label{IIB} 

In spite of the good theoretical foundation behind meson-exchange Feynman amplitudes, meson theory does not 
 provide a systematic approach to constructing nuclear forces. As mentioned in the Introduction, chiral EFT presents 
the opportunity for such systematic development.

Crucial for a nuclear EFT are the processes of regularization and renormalization.
Concerning the former, 
all chiral interactions are 
multiplied by a regulator function
which typically has the form:
\begin{equation}
f(p',p) = \exp[-(p'/\Lambda)^{2n} - (p/\Lambda)^{2n}] \; ,
\label{reg}
\end{equation}
where $\Lambda$ is known as the cutoff parameter.

 Nucleon-nucleon potentials have been developed at different orders and cutoff 
values~\cite{EM03,chinn5}.  
 Chiral EFT predictions allow for the quantification of uncertainties that stem from the              
truncation error and cutoff variations (as well as additional 
sources of errors).

\begin{figure}[!t] 
\centering         
\vspace*{1.2cm}
\hspace*{0.3cm}
\scalebox{0.4}{\includegraphics{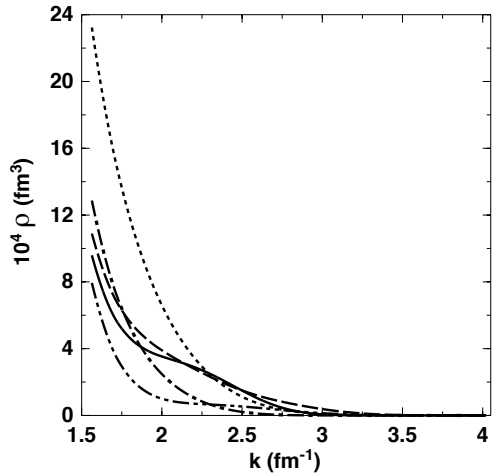}}
\vspace*{-2.5cm}
\caption{(Color online)                                                                         
Momentum distributions in the deuteron predicted with chiral potentials at: LO (dotted);  
NLO (dash-double dot); N$^2$LO (dash-dot); N$^3$LO (dash); and N$^4$LO (solid). 
The cutoff is fixed at 500 MeV.                           
} 
\label{FIG2}
\end{figure}

Consistent with that philosophy, in Fig.~\ref{FIG2} we show the momentum distribution in the deuteron, 
including five orders of the chiral expansion. We note that the potential at N$^4$LO is 
a preliminary version of 
a high-precision nucleon-nucleon potential at fifth order~\cite{n4lo}. 
(Some wavy structures noticeable in the figure are most likely due to the 
polynomial nature of the EFT contacts.)
We observe huge variations at the lowest orders, particularly from LO to NLO, 
and a clear convergence pattern with increasing order.

\begin{table}                
\centering \caption                                                    
{Probabilities of SRC and $D$-state probabilities for the chiral interactions considered in Fig.~\ref{FIG2}.            
} 
\vspace{5mm}
\begin{tabular}{|c|c|c|}
\hline
Model & $a_{2N}(d)$ & $P_D$ \\
\hline     
\hline     
 LO &  0.047& 0.0757  \\
    &   &  \\ 
 NLO &  0.015& 0.0313  \\
   &    &   \\ 
 N$^2$LO &  0.022& 0.0417  \\
   &    &    \\ 
 N$^3$LO &  0.030& 0.0451  \\
  &    &    \\ 
 N$^4$LO &  0.026& 0.0414  \\
\hline
\end{tabular}
\label{tab2} 
\end{table}

 Table~\ref{tab2} shows the integrated probabilities corresponding to Fig.~\ref{FIG2}.       
At the higher, converging orders, the SRC probabilities are not very different from the one 
predicted by CD-Bonn.
That is, 
chiral potentials with suitable cutoff can be constructed with excellent fit to the NN data and 
off-shell nature similar to the highest-quality {\it non-local} meson-exchange forces. 

The truncation error at order $n$ is defined as the difference between the predictions at orders $n+1$ and $n$. Thus, the error of $a_{2N}(d)$ at N$^3$LO is $\pm$0.004.
We have also considered variations of the cutoff parameter between 
500 and 600 MeV.                                                            
At N$^3$LO, we found the value of $a_{2N}(d)$ with cutoff of 600 MeV to be nearly the same as with 500 MeV. Therefore, cutoff uncertainty is below the truncation error, and our final result at N$^3$LO can be stated as 0.030$\pm$0.004.
Concerning the uncertainty of the N$^4$LO result, the prediction for the next higher order is unknown. Therefore, assuming (pessimistically) the same 
truncation error as at N$^3$LO, the prediction at 
 N$^4$LO can be stated as 0.026$\pm$0.004.

\subsection{Discussion} 
\label{IIC} 

The deuteron is the simplest system where off-shell behavior can be explored. Characteristic differences
exist between meson-theoretic potentials using 
fully relativistic one-pion exchange amplitudes (that is, non-local tensor forces) and those which use 
static one-pion-exchange.                        
{\it Off-shell behavior is not observable and thus cannot be uniquely determined by measurements.} The best one can do is
to have a good theoretical foundation for it. In meson theory, this is provided by relativistic meson-exchange amplitudes.

In Section~\ref{IIA} we saw that the                      
SRC probability  predicted with CD-Bonn is roughly 25\% below the value of $a_{2N}(d)$ 
 cited in Ref.~\cite{CLAS} and used to evaluate absolute probabilities 
in heavier nuclei.       
A discrepancy of qualitatively similar nature                   
exists for the wound integral in nuclear matter. More precisely, conventional non-local potentials are 
known to predict about 10-15\% for the wound integral (at normal density)~\cite{FS14}, whereas a value of about 25\% is cited from 
 extrapolation to nuclear matter of the empirical information~\cite{Pia13}. In fact, the ratio of the probability for   
nucleus $A$ to the deuteron, extrapolated to infinite symmetric matter, is given to be 6.5$\pm$1.0~\cite{Pia13}.
With the absolute probability for the deuteron taken equal to 0.04, the value cited above is obtained. 
However, when the $a_{2N}(d)$=0.032 value from CD-Bonn is used, one obtains 20\%, and with 
$a_{2N}(d)$=0.026 as from the converged chiral results, a value of 17\% is 
obtained for empirical short-range correlations in nuclear matter, which is getting closer to the predicted
wound integrals.

In summary, 
the question of consistency between description of short and long range physics seems to go beyond the
(intrinsically) 
``low-momentum" nature of some potentials. Instead, it points to 
non-locality in the tensor force, a feature which has been found since a long time to be very attractive 
in nuclear structure. 
So, we are revisiting an old issue which resurfaces 
in the light of new experiments. We suggest to take a broad view of it, combining our former
and present understanding of microscopic nuclear forces and their development.

\section{Conclusions and outlook}                                                                  
\label{Concl} 

The deuteron is a beautifully simple benchmark for theories of nuclear forces.
From this study, we conclude that predictions of high-momentum distributions in the deuteron 
with high-quality non-local meson-exchange forces or state-of-the-art chiral forces 
are systematically lower than what is used to extract empirical information for heavier nuclei.  
Taking those results into account leads to a better agreement between SRC in nuclear matter and 
theoretical predictions. 

We plan to extend our microscopic analysis to the A=3 system using a broad spectrum of interactions as in the present
study. We hope this will shed more light on how to reconcile theory and empirical analyses. 

Finally,                    
some caution should be excercised in the interpretation of the empirical information discussed above 
as an experimental constraint on the off-shell behavior. 

\section*{Acknowledgments}
This work was supported by 
the U.S. Department of Energy, Office of Science, Office of Basic Energy Sciences, under Award Number DE-FG02-03ER41270. 

\end{document}